\pdfoutput=1

\documentclass[aps,prb,showpacs,floatfix,twocolumn,letterpaper,longbibliography,superscriptaddress,notitlepage]{revtex4-1}
\usepackage{graphicx}
\usepackage[english]{babel}
\usepackage{bm}
\usepackage{amsmath,amsfonts}
\usepackage{amsbsy}
\usepackage[colorlinks=true,linkcolor=blue,urlcolor=blue,citecolor=blue,breaklinks=true]{hyperref}
\usepackage[utf8]{inputenc}
\DeclareUnicodeCharacter{03BC}{\mbox{$\mu$}}
\DeclareUnicodeCharacter{2009}{ }

\usepackage{graphics}
\usepackage{xcolor}
\usepackage{bbm}
\def\bm#1{\mbox{\boldmath{$#1$}}}
\def\rr#1{(\ref{#1})}
\newcommand{\be}{\begin{equation}}
\newcommand{\ee}{\end{equation}}
\def\boldsymbol#1{\mbox{\boldmath$#1$}}
\usepackage[normalem]{ulem}
\usepackage{graphics}
\usepackage{amsmath}
\usepackage{amsfonts}
\usepackage{bm}
\usepackage{color}
\usepackage{bbm}
\def\bm#1{\mbox{\boldmath{$#1$}}}
\def\rr#1{(\ref{#1})}
\def\boldsymbol#1{\mbox{\boldmath$#1$}}
\newcommand{\Str}{\mbox{str}}\newcommand{\tr}{\mbox{tr}}
\newcommand{\bphi}{\mbox{\boldmath $\phi$}}
\newcommand{\bvarphi}{\mbox{\boldmath $\varphi$}}
\newcommand{\bPsi}{\mbox{\boldmath $\Psi$}}
\newcommand{\bpsi}{\mbox{\boldmath $\psi$}}
\newcommand{\nnabla}{\mathbf \nabla}
\newcommand{\ptau}{\partial_\tau}
\newcommand{\bs}{\mbox{\boldmath $\sigma$}}
\newcommand{\req}[1]{Eq.~(\ref{#1})}
\newcommand{\reqs}[1]{Eqs.~(\ref{#1})}
\newcommand{\rref}[1]{(\ref{#1})}
\newcommand{\h}{\mathbf{h}}
\renewcommand{\S}{\mathbf{S}}
\newcommand{\w}{\omega}
\newcommand{\W}{\Omega}
\newcommand{\q}{\mathbf{q}}\newcommand{\A}{\mathbf{A}}
\newcommand{\p}{\mathbf{p}}
\renewcommand{\P}{\mathbf{P}}
\newcommand{\n}{\mathbf{n}}\newcommand{\N}{\mathbf{N}}
\renewcommand{\r}{\mathbf{r}}
\renewcommand{\k}{\mathbf{k}}
\newcommand{\bchi}{\mbox{\boldmath $\chi$}}
\newcommand{\M}{\mathbf{M}}
\renewcommand{\a}{\mathbf{a}}\renewcommand{\v}{\mathbf{v}}
\newcommand{\pt}{\partial_t}

\newcommand{\beq}{\begin{equation}}
\newcommand{\eeq}{\end{equation}}
\newcommand{\beqa}{\begin{eqnarray}}
\newcommand{\eeqa}{\end{eqnarray}}
\newcommand{\bea}{\begin{eqnarray}}
\newcommand{\eea}{\end{eqnarray}}
\usepackage{amssymb}
\usepackage{array}
\usepackage{amsmath}
\usepackage{graphicx}\usepackage{graphics}
\usepackage{dcolumn}
\usepackage{bm}\usepackage{varioref}
\newcommand{\hH}{\hat{H}}\newcommand{\hA}{\hat{A}}\newcommand{\hF}{\hat{F}}

\newcommand{\hG}{\hat{\mathcal G}} \newcommand{\hbarG}{\hat{\bar{\mathcal G}}}
\newcommand{\hLambda}{\hat{\Lambda}}
\newcommand{\hSigma}{\hat{\Sigma}}
\newcommand{\hP}{\hat{P}} \newcommand{\hbarP}{\hat{\bar{P}}}
\newcommand{\htau}{\hat{\tau}}

\newcommand{\hJ}{\hat{\mathbb J}}
\newcommand{\hI}{\hat{\mathbb I}}
\newcommand{\hK}{\hat{\mathbb K}}
\newcommand{\hHO}{\hat{\mathbb H}^{(1)}}
\newcommand{\hHT}{\hat{\mathbb H}^{(2)}}
\newcommand{\hHOT}{\hat{\mathbb H}^{(1,2)}}
\newcommand{\R}{\mathbf R}\newcommand{\B}{\mathbf B}

\newcommand{\hbarJ}{\hat{\bar{\mathbb J}}}
\newcommand{\hbarI}{\hat{\bar{ \mathbb I}}}
\newcommand{\hbarK}{\hat{\bar{ \mathbb K}}}
\newcommand{\hbarHO}{\hat{\bar{\mathbb H}}^{(1)}}

\newcommand{\hbarHT}{\hat{\bar{\mathbb H}}^{(2)}}
\newcommand{\hbarHOT}{\hat{\bar{\mathbb H}}^{(1,2)}}

\newcommand{\sgn}{{\text{ sgn }}}
\renewcommand{\b}{{\mathbf b}}\renewcommand{\t}{{\mathbf t}}
\newcommand{\cE}{{\cal E}}
\usepackage{physics}

\newcommand{\tin}{\tau_{\mathrm{in}}}
\newcommand{\tel}{\tau_{\mathrm{el}}}
\usepackage{subcaption}
\usepackage{float}

\newcommand{\bps}{\bm p_{\mathrm s}}
\newcommand{\hatGd}{\hat G(i\tilde\epsilon_n, \bm p)}
\newcommand{\hatGdagd}{\hat G^\dagger(i\tilde\epsilon_n, \bm p)}

\usepackage{float}

\DeclareFontFamily{U}{mathx}{\hyphenchar\font45}
\DeclareFontShape{U}{mathx}{m}{n}{
      <5> <6> <7> <8> <9> <10>
      <10.95> <12> <14.4> <17.28> <20.74> <24.88>
      mathx10
      }{}
\DeclareSymbolFont{mathx}{U}{mathx}{m}{n}
\DeclareFontSubstitution{U}{mathx}{m}{n}
\DeclareMathAccent{\widecheck}{0}{mathx}{"71}
\DeclareMathAccent{\wideparen}{0}{mathx}{"75}

\begin{document}

\title{Giant magnetoconductivity in non-centrosymmetric superconductors}

\author{M. Smith}

\affiliation{Department of Physics, University of Washington, Seattle, WA 98195 USA}

\author{A. V. Andreev}

\affiliation{Skolkovo Institute of Science  and  Technology,  Moscow,  143026,  Russia}

\affiliation{Department of Physics, University of Washington, Seattle, WA 98195 USA}

\affiliation{L. D. Landau Institute for Theoretical Physics, Moscow, 119334 Russia}

\author{B. Z. Spivak}

\affiliation{Department of Physics, University of Washington, Seattle, WA 98195 USA}

\date{\today}

\begin{abstract}
We discuss a novel physical mechanism which gives rise to a giant magnetoconductivity in non-centrosymmetric superconducting films. This mechanism is caused by a combination of spin-orbit interaction and inversion symmetry breaking in the system, and arises in the presence of an in-plane magnetic field ${\bf H}_\|$. It produces a contribution to the conductivity, which displays a strong dependence on the angle between the electric field ${\bf E}$ and ${\bf H}_\|$, and is proportional to the  inelastic relaxation time   of quasiparticles. Since in  typical situations the latter is much larger than the elastic one this contribution can be much larger than the conventional conductivity thus leading to giant microwave absorption.
\end{abstract}

\maketitle

\addtolength{\abovedisplayskip}{-1mm}

Non-centrosymmetric conductors with spin-orbit interaction exhibit interesting phenomena which are forbidden in centrosymmetric materials.  Examples include spin magnetization induced by a current in metals~\cite{dyakonov_possibility_1971, dyakonov_current-induced_1971,levitov_magnetoelectric_1985, dyakonov_spin_2017}, the quantized anomalous Hall effect on the surface of topological superconductors~\cite{hasan_colloquium_2010},   linear coupling between supercurrent and an in-plane magnetic field in superconducting films~\cite{levitov_magnetostatics_1985, edelstein_magnetoelectric_1995,yip_two-dimensional_2002,edelstein_magnetoelectric_2005,smidman_superconductivity_2017},
 and the superconducting diode effect~\cite{buzdin_direct_2008,szombati_josephson_2016,hoshino_nonreciprocal_2018,baumgartner_josephson_2021,yuan_supercurrent_2021}.
In this article we consider a new effect of this kind; we show that non-centrosymmetric superconducting films subjected to an in-plane magnetic field ${\bf H}_{\|}$ exhibit a giant anisotropic dissipative  \emph{ac} conductivity $\sigma$, which can be measured in microwave absorption experiments.

The \emph{ac}  conductivity $\sigma$  relates the current density $\bm{j}$ to the applied  electric field $\bm{E}(t)=\bm{E}_{0}\cos \omega t$.
At sufficiently low frequency $\omega$ this relation takes  the form
\begin{equation}\label{eq:conductivity_def}
  \bm{j} = \bm{j}_{\mathrm{eq}} (\bm{p}_{\mathrm s} )  + \hat{\sigma} \bm{E}(t),
\end{equation}
where  $\hat{\sigma}$ is a conductivity tensor, and  $\bm{j}_{\mathrm{eq}} (\bm{p}_s )$ is the equilibrium current evaluated at the instantaneous value of the superfluid momentum $\bm{p}_{\mathrm s}$. The latter is defined as
$\bm{p}_{\mathrm s} = \frac{1}{2}\left(\bm{\nabla} \chi - \frac{2e}{c}\bm{A}\right)$, where $\chi$ is the order parameter phase,  $\bm{A}$ is the vector potential, and $c$ is the speed of light. In the presence of  microwave radiation $\bm{p}_{\mathrm s}$  evolves in time according to
\begin{equation}\label{eq:E-Ps}
\frac{d \bm{ p}_{s}}{d t} =   e{\bf E}(t)  .
\end{equation}
The first term  in the right hand side of Eq.~\eqref{eq:conductivity_def} represents the dissipationless supercurrent. In the presence of superfluid momentum and in-plane magnetic field $\bm{H}_{\|}$  it can be written in the form
 \begin{equation}\label{eq:supercurrent}
\bm{j}(\bm{p}_s)=\frac{e}{m}N_{s}\bm{p}_{s} + a{\bf [n\cross h]}+b{\bf h}.
\end{equation}
Here $e$ and $m$ denote the electron charge and mass,  $N_{s}$ is the superfluid density, and we introduced the notation ${\bf h} = g\mu_{\mathrm B} \bm{H}_{\|}/2$, with $g$ being the electron $g$-factor, and $\mu_{\mathrm B}$ the Bohr magneton.
The last two terms in the right hand side of Eq.~\eqref{eq:supercurrent} describe the magnetoelectric effect~\cite{edelstein_magnetoelectric_2005} and   are allowed by symmetry  only in non-centrosymmetric superconductors.

We will assume that the electric field ${\bf E}$ is applied in the $x$-direction. In this case the  component $\sigma_{xx}=\sigma$ of the conductivity tensor
$\hat{\sigma} $ is responsible for the microwave absorption.
The dissipative part of the \emph{ac} conductivity of  centrosymmetric superconductors has been extensively studied starting with the classic work of Bardeen and Mattis~\cite{mattis_theory_1958}.  In the absence of a \emph{dc} supercurrent
 and at sufficiently low frequencies, the value of the conductivity  $\sigma$  is controlled by elastic scattering of  quasiparticles off impurities and is proportional to the  quasiparticle
 elastic  momentum relaxation time.  In particular, in the vicinity of the critical temperature $\sigma$ is nearly equal to the normal state conductivity   $\sigma_{\mathrm{n}}=e^{2} \nu_{\mathrm{n}} D$, where $\nu_{\mathrm{n}} $  is the density of states at the Fermi level, and $D=v_{\mathrm F}^{2}\tel/3$  is the diffusion coefficient, with $v_{F}$ being the Fermi velocity and $\tel$ the elastic relaxation time in the normal state.

Below, we show that for non-centrosymmetric superconductors placed in an in-plane magnetic field $\bm{H}_{\|}$ the linear conductivity  $\sigma$  acquires an additional contribution $\sigma_{\mathrm{DB}} $, which is proportional to the inelastic relaxation time of quasiparticles, $\tin$, and has a pronounced dependence on the angle between the microwave field $\bm{E}$ and $\bm{H}_{\|}$.  Since in typical superconductors $\tin$ exceeds $\tel$ by several orders of magnitude this leads to a giant anisotropic  magnetoconductivity.  The frequency dispersion of $\sigma_{\mathrm{DB}}(\omega)$ takes place at relatively low frequency $\omega\sim 1/\tin$,  which is much smaller than the characteristic dispersion frequency $1/\tel$ of the normal state conductivity  $\sigma_{\mathrm{n}}$.

The existence of this phenomenon can be traced to the fact that the quasiparticle density of states in superconductors   $\nu(\epsilon, {\bf p}_{\mathrm s})$ depends on the superfluid momentum $\bm{p}_{\mathrm s}$.  In the presence of microwave radiation the latter changes with time according to Eq.~\eqref{eq:E-Ps} producing a time-dependent density of states.
At small frequencies this may be described in terms of the spectral flow,  i.e. motion of individual  quasiparticle energy levels in  energy space. The quasiparticles which occupy these levels are entrained by the spectral flow.
As a result, the quasiparticle distribution acquires a non-equilibrium component.
Its relaxation causes energy dissipation at a rate that is proportional to the inelastic relaxation time $\tin$. This dissipation mechanism is similar to the Debye mechanism in centrosymmetric superconductors~\cite{smith_giant_2020,smith_debye_2020,smith_conductivity_2020}.  We therefore refer to it as the Debye contribution to conductivity and denote it by $\sigma_{\mathrm{DB}}$.  The total dissipative part of the conductivity is given by the sum of $\sigma_{\mathrm{DB}}$ and the conventional contribution proportional to $\tau_{\mathrm{el}}$.

Due to the scalar character of the density of states   $\nu (\epsilon, {\bf p}_{\mathrm s})$  its  linear coupling to ${\bf p}_{\mathrm s}$ induced by the microwave field $\bm{E}$ is possible only  if  the symmetry of the system allows for a presence of a polar vector. Therefore, in centrosymmetric superconductors the contribution proportional to $\tau_{\mathrm{in}}$ is possible only in a current-carrying state~\cite{ovchinnikov_electromagnetic_1978,smith_debye_2020}. In superconductors with broken inversion symmetry a polar vector  enabling  linear coupling of the density of states to microwave radiation may be formed from the pseudovector $\bm{H}_{\|}$.  As a result, in films of non-centrosymmetric superconductors placed in an in-plane magnetic field, the linear conductivity acquires a contribution proportional to $\tau_{\mathrm{in}}$ even  in the absence of \emph{dc} supercurrent.

We note that the presence of such an additive contribution to the conductivity is in drastic contrast with the phenomenological Matthiessen's rule, which states that the
resistivity (including the microwave resistivity of superconductors) is proportional to the sum of partial momentum relaxation rates due to each type of relaxation process. The latter implies that the resistivity  is controlled by the shortest relaxation time.

We begin our quantitative treatment by writing down the Hamiltonian of a normal metal,
\begin{equation}
\label{eq:Hamiltonian}
	H_n ({\bf  p})= \xi_{\bf{p}}  -  \bm{b}(\bm{p})\cdot\bm\sigma - {\bf h}\cdot\bm\sigma,
\end{equation}
where  $\xi_{\bf{p}} = E(\bm{p}) - E_F$ is the quasiparticle energy relative to the Fremi energy $E_{F}$. Below we assume isotropic dispersion, $E(\bm{p}) =p^2/2m$, and the spin-orbit coupling of the form
\begin{equation}\label{eq:so}
  \bm{b}(\bm{p}) = \alpha \bm{p}\times\bm{n} + \beta \bm{p},
\end{equation}
which includes both Rashba ($\beta =0$) and Deresselhaus ($\alpha =0$) cases.
Here  $\bm{\sigma}$ are spin Pauli matrices, and  $\bm{n}$ is one of the two non-equivalent normals to the sample.

The eigenstates of the Hamiltonian \eqref{eq:Hamiltonian} can be characterized by helicity (projection of spin on the direction of $\bm{b}(\bm{p}) + {\bm h}$).  The dispersion of the bands with positive and negative helicity has the form
\begin{align}
\label{eq:nmspectrum}
\epsilon_{\bf p\pm}= \xi_{\bf{p}} \pm |\bm b(\bm p) + \bm h|.
\end{align}
 Although the spectrum Eq.~\eqref{eq:nmspectrum} contains an odd in ${\bf p}$ component, the equilibrium current in the normal state vanishes;  ${\bf j}_{N}=
\sum_{\pm}\int n_{\mathrm F} (\epsilon_{\bf p\pm}) d {\bf p}=0$.  Here $n_{\mathrm F}$ is the Fermi distribution function.

In non-centrosymmetric superconductors in the presence of an in-plane  magnetic field the equilibrium current may be written in the form of Eq.~\eqref{eq:supercurrent}. To evaluate the coefficients $a,b, N_{\mathrm s}$ one can use the BCS Hamiltonian
for superconductors with spin-orbit coupling. We write our Hamiltonian as  $H = H_0 + U(\bm r) \tau_3$, where
\begin{eqnarray}
\label{eq:HSC}
	H_0 (\bm{p}) &=& \left(\begin{array}{cc}
		H_n(\bm{p}) & \hat\Delta\\
		\hat\Delta & -H_n^T (-\bm{p})
	\end{array} \right)     \nonumber   \\
	&=& \left(\xi_{\bf{p}} - \bm{b}(\bm{p})\cdot\bm\sigma\right)\tau_3 + \hat\Delta \tau_1 - \bm h\cdot\bm\sigma   .  
\end{eqnarray}
Here $\bm\tau$ are the Pauli matrices in the  Gorkov-Nambu space, $U(\bm r)$ is the random impurity potential, whose strength will be characterised by the value of the electron elastic mean free time $\tel$.   For simplicity, we assume local interactions. In this case the order parameter is a singlet, and $\hat{\Delta} = \Delta \hat{I}$  is proportional to the identity matrix in spin space. The superfluid momentum is included by making the gauge transformation $\bm p \rightarrow \bm p + \bm p_{\mathrm s} \tau_3$ .

 We focus on the temperature interval near  the critical temperature, $T_{c}-T\ll T_{c}$. In this  case the coefficients $a$ and $b$  in Eq.~\eqref{eq:supercurrent} can be evaluated in the second order   in  $\Delta$ using the conventional diagram technique,
\begin{equation}\label{eq:ab}
  a =  e\alpha m \frac{\Delta^2}{  T_{c}^2}  g (T_c \tau_{\mathrm{el}}) , \quad b =  e \beta m \frac{\Delta^2}{T_{c}^2} g (T_c \tau_{\mathrm{el}}),
\end{equation}
 In  the clean and dirty cases the function $g (T_c \tau_{\mathrm{el}})$   is given by
\begin{equation}\label{eq:g_cases}
   g(T_c \tau_{\mathrm{el}}) = \begin{cases}  \frac{7}{8 \pi^2} \zeta(3) , &  T_{c}\tel \gg 1, \\
  \frac{3}{2 \pi^2} \zeta(2)  \, T_{c}\tel, &  T_{c} \tel \ll 1,
    \end{cases}
\end{equation}
where $\zeta(n)$ is the Reimann  zeta function, and we have assumed $\alpha p_F, \beta p_F \gg \max\left\{T, \sqrt{T/\tel} \right\}$.

 In the ground state the current density must vanish. Substituting Eqs.~\eqref{eq:ab}, \eqref{eq:g_cases} into Eq.~\eqref{eq:supercurrent}
we obtain  the value ${\bf p}_{\mathrm s}^{(\mathrm{gs})}$ of ground state superfluid momentum in  the form
\begin{equation}\label{eq:GS_ps}
  {\bf p}_{\mathrm s}^{(\mathrm{gs})} = \tilde a \bm n \times \bm h  + \tilde b  \bm h,
\end{equation}
where the coefficients $\tilde a$ and $\tilde b$ are given by
\begin{subequations}\label{eq:ps_cases}
  \begin{align}
\label{eq:tilde_ab}
\tilde a & =\frac{m\alpha}{2 E_F}  \tilde{g} (T_c\tel), \quad \tilde b  =\frac{m\beta}{2 E_F}  \tilde{g} (T_c\tel),
 \\
\tilde{g} (T_c\tel) & =  \begin{cases}
	1, \quad T_{c}\tel \gg 1,\\
	2 \quad T_{c}\tel \ll 1.
\end{cases}
\end{align}
\end{subequations}
In deriving Eq.~\eqref{eq:ps_cases} we have used the standard results~\cite{abrikosov_methods_1975}
\begin{align}
\label{eq:GS_Ns}
N_s &=  \begin{cases}
\frac{7 p_F^2}{8\pi^3} \zeta(3) \frac{\Delta^2}{T^2}	, \quad T_c\tel \gg 1\\
\frac{p_F^2}{8} \frac{\Delta^2\tel}{T},	 \quad T_c\tel \ll 1
\end{cases}
\end{align}
 for the superfluid density.
Equations \eqref{eq:ab} - \eqref{eq:ps_cases} were obtained by Edelstein~\cite{edelstein_magnetoelectric_1995,edelstein_magnetoelectric_2005} for the case of Rashba spin-orbit coupling. For a more general spin-orbit coupling the derivation is presented in  Appendix A.

We now turn to the consideration of the dissipative part of the conductivity.
A general expression for the Debye contribution to the linear microwave conductivity,  $\sigma_{\mathrm{DB}}(\omega)$  was  obtained in Ref.~\onlinecite{smith_debye_2020}.  For finite microwave frequencies $\omega$  it has the form
\begin{equation}
\label{eq:sigma_ratio}
    \frac{\sigma_{\mathrm{DB}}}{\sigma_{\mathrm n}} = \frac{3}{4} \frac{\tau_{\mathrm{in}}}{\tau_{\mathrm{el}}} \frac{1}{1+(\omega \tin)^{2}}\int \frac{d\epsilon}{T} \frac{\nu(\epsilon)}{\nu_{\mathrm n}} \frac{V^2(\epsilon)}{v_{\mathrm{F}}^{2} \cosh \left(\epsilon/(2T)\right)},
\end{equation}
where
\begin{equation}
\label{eq:LevelSensitivity2}
	\bm{V}(\epsilon) =- \frac{1}{\nu(\epsilon)} \int_0^\epsilon d\tilde\epsilon \frac{\partial \nu(\epsilon)}{\partial \bm{p}_{\mathrm s}}
\end{equation}
characterizes the sensitivity of quasiparticle energy levels to changes in $\bm{p}_{\mathrm s}$.
The derivation of Eqs.~\eqref{eq:sigma_ratio} and \eqref{eq:LevelSensitivity2} is based on the concept of spectral flow.
Accordingly, the value of $\sigma_{DB}$ is completely determined  by the dependence of the quasiparticle density of states in a superconductor, $\nu(\epsilon, {\bf p}_{s})$, on the superfluid momentum.

In $s$-wave superconductors  $\sigma_{\mathrm{DB}}$  given by   the integral in Eq.~\eqref{eq:sigma_ratio} is controlled by the energy interval close to the gap $\epsilon-\Delta\ll \Delta$.
The derivative in Eq.~\eqref{eq:LevelSensitivity2} must be evaluated at zero current, that is at ${\bf p}_{\mathrm s} = {\bf p}_{\mathrm s}^{(\mathrm{gs})}$, where ${\bf p}_{\mathrm s}^{(\mathrm{gs})}$   is given by Eq.~\eqref{eq:GS_ps}. In the absence of time-reversal symmetry breaking, $h= p^{(gs)}_{\mathrm{s}} =0$, the density of states has the BCS square-root singularity at $\epsilon =\Delta$. At $h\neq 0$  this singularity is broadened.  The Debye conductivity $\sigma_{\mathrm{DB}}$ is dominated by the motion of quasiparticle energy levels inside this energy interval. Its character depends on the magnitude of the magnetic field $h$ and the elastic relaxation rate in the normal state, $1/\tel$.

One can distinguish between  ballistic and diffusive regimes of microwave absorption. In the former, the broadening exceeds the rate of elastic scattering of quasiparticles, whose energies lie inside the broadened BCS singularity. In the latter the opposite inequality takes place.  The value of the magnetic field $h$ separating the ballistic and diffusive regimes  may be estimated by equating $h$   with the elastic relaxation rate for quasiparticles in the energy interval  $|\epsilon-\Delta|\sim h$.  Recalling that elastic relaxation rate
 for quasiparticles depends on the energy $\epsilon$  as (see, for example, Ref.~\onlinecite{mineev_introduction_1999})
\[
	\frac{1}{\tau^{\mathrm s}_{\mathrm{el}}(\epsilon) }\sim    \frac{1}{\tel}\sqrt{\frac{\epsilon-\Delta}{\Delta}}
\]
one finds that the crossover between the ballistic and the diffusive regimes  occurs at
$ h \tau_{\mathrm{el}}^2\Delta  \sim 1$.

\emph{Ballistic regime},  $h \tau_{\mathrm{el}}^2\Delta \gg 1$:
 In the superconducting state,  the excitation spectrum of  quasiparticles with positive and negative helicity in the presence of an in-plane Zeeman field ${\bf H}_{\|}$ and $\bm{p}_{\mathrm{s}}$  takes the form (up to first order in $h\ll |{\bf b}|$ and $p_{\mathrm s}$)
\begin{align}
	\tilde\epsilon_\pm (\bm{p}) = \sqrt{(\xi_{\bf{p}} \pm |\bm b(\bm p)|)^2  + \Delta^2}  + \bm{\tilde v}_\pm (\bm{p})\cdot\bm p_{\mathrm s} \mp \bm{h} \cdot  \bm{\hat{b}}  (\bm{p}) . \nonumber
\end{align}
Here  $\bm{\hat{b}} = \bm b(\bm p) /|\bm b (\bm p)|$, and $\bm{\tilde v}_\pm (\bm{p}) = \bm{p}/m \pm \bm{b}(\bm{p}) /|\bm{p}|$  is the normal state  velocity in the band with helicity $\pm$. The resulting density of states is given by
\begin{align}
\label{eq:CleanDOS}
\nu(\epsilon) = &  \sum_{\pm} \int\frac{d\phi}{4 \pi} \frac{ \nu_\pm \left( \epsilon - \bm{\tilde{v}}_{\mathrm{F}\pm} \cdot \bm{p}_{\mathrm s} \pm \bm{h} \cdot\bm{\hat{b}} (\bm{p})  \right) }{\sqrt{ [\epsilon - \bm{\tilde{v}}_{\mathrm{F}\pm} \cdot \bm{p}_{\mathrm s} \pm \bm h \cdot\bm{\hat{b}} (\bm{p}) ]^2 -\Delta^2} },
\end{align}
 where $\bm{\tilde{v}}_{\mathrm{F}\pm}$ is the band velocity, $\bm{\tilde v}_{\pm} (\bm{p}) $ evaluated on the  Fermi circle with the corresponding helicity ($\pm$),  $\phi$ is the azimuthal angle of $\bm{p}$,
and $\nu_\pm = \nu_{\mathrm{n}}  \left(1 \pm b/E_F\right)$ is the density of states on the corresponding Fermi circle.

The superfluid momentum in Eq.~\eqref{eq:CleanDOS} can be written  as
$\bm{p}_{\mathrm{s}}={\bf p}^{\mathrm{(gs)}}_{\mathrm s}+\delta {\bf p}_{ \mathrm s}$, where $\delta {\bf p}_{\mathrm s}(t)$ is the superfluid  momentum  related to the electric field  by Eq.~\eqref{eq:E-Ps}. For the  case where the spin-orbit coupling has the form of Eq.~\eqref{eq:so}, to linear order in $\delta {\bf p}_{ \mathrm s}$, the density of states in Eq.~\eqref{eq:CleanDOS} depends only on the component of $\delta {\bf p}_{ \mathrm s}$ that is parallel to ${\bf p}^{\mathrm{(gs)}}_{\mathrm s}$.   This follows from the fact that, according to  Eqs.~\eqref{eq:so},   \eqref{eq:GS_ps}, and \eqref{eq:tilde_ab} for $ \bm{p}_{\mathrm s} = {\bf p}^{\mathrm{(gs)}}_{\mathrm s}$  the anisotropic terms $ \bm{\tilde{v}}_{\mathrm{F}\pm} \cdot \bm{p}_{\mathrm s} $ and $\bm{h} \cdot\bm{\hat{b}} (\bm{p})  $ in Eq.~\eqref{eq:CleanDOS} have identical dependence on the azimuthal angle $\phi$.  Assuming  the longitudinal polarization of the electric field, $\bm{p}_{\mathrm s} \parallel  {\bf p}^{\mathrm{(gs)}}_{\mathrm s}$, and performing the angular integration in Eq.~\eqref{eq:CleanDOS} we get the density of states in the form
\begin{equation}
\label{eq:CleanDOS_2}
	\nu(\epsilon) =\frac{1}{2\pi} \sum_{\pm}  \nu_{\pm} \gamma_\pm^{-1/2} \, \theta\left(w+\gamma_\pm\right)  \Re  \, K\left(\frac{w+\gamma_\pm}{2\gamma_\pm}\right).
\end{equation}
Here  $K(x) = \int_0^{\pi/2}d\phi(1-x\sin^2\phi)^{-1/2}$ is the complete elliptic integral of the first kind, and we introduced the following variables:  $w = (\epsilon-\Delta)/\Delta\ll 1$, and $\gamma_\pm =\left( h \pm v_{\mathrm F}p^{(gs)}_{\mathrm s}\right) /\Delta$.

 Next, we evaluate the level sensitivity,  Eq.~\eqref{eq:LevelSensitivity2}. There are two energy intervals to consider. For energies $-\gamma_-\leq w$ the bands with opposite helicities give almost equal but opposite contributions to $V$, and the level sensitivity is $V \sim v_F b/E_F$. Therefore, the contribution of this interval  to the conductivity in Eq.~\eqref{eq:sigma_ratio} is quadratic in spin-orbit coupling strength,
$\propto \left(b/E_F\right)^2$. Inside the second energy interval, $-\gamma_+ \leq w \leq -\gamma_-$,  only the band with $+$ helicity contributes to the density of states
 in  Eq.~\eqref{eq:CleanDOS_2}. As a result, the level  sensitivity in this energy interval is given by
\begin{equation}
\label{eq:CleanSLevelSensitivity}
V(\epsilon) = - v_{\mathrm F}.
\end{equation}
 Therefore, although the  width of this interval is relatively small, $2 v_{\mathrm F} p_{\mathrm s}^{(\mathrm{gs})} \propto b/E_{\mathrm F} $,  it provides the main contribution to the conductivity in Eq.~\eqref{eq:sigma_ratio}.
Substituting  Eq.~\eqref{eq:CleanSLevelSensitivity} into Eq.~\eqref{eq:sigma_ratio}, and performing integration over $\epsilon$  in the interval  $-\gamma_+ \leq w \leq -\gamma_-$ we obtain the Debye contribution to the longitudinal conductivity, ($\bm{E}_0 \parallel  {\bf p}^{\mathrm{(gs)}}_{\mathrm s}$). The angular dependence of the Debye conductivity relative to the orientation of magnetic field is then restored with the aid of Eqs.~\eqref{eq:GS_ps} and \eqref{eq:tilde_ab},
\begin{equation}
\label{eq:sigma_ration_ballistic}
	\frac{\sigma_{\mathrm{DB}}}{\sigma_{\mathrm n}} =  \frac{3}{16\left(1+(\omega\tin)^2\right)}   \frac{\tau_{\mathrm{in}}}{\tau_{\mathrm{el}}}    \frac{p_F \sqrt{\Delta h} }{T E_F } \,  \frac{\left(\beta\cos\theta - \alpha\sin\theta\right)^2 }{\sqrt{\alpha^2+\beta^2}}.
\end{equation}
Here $\theta$ is the angle between the electric field $\bm E$ and the in-plane magnetic field $\bm h$.

\emph{Diffusive regime},  $h \tau_{\mathrm{el}}^2\Delta \ll1$: In this case we express the single particle density  of states in terms of the retarded single-particle Green's function
\begin{align}
\label{eq:DOS_GF}
	\nu(\epsilon) = -\frac{1}{\pi}\Im \int\frac{d^2 p}{(2\pi)^2} \tr \hat{G}^R(\bm{p},\epsilon)
\end{align}
where the trace is performed over both  Gorkov-Nambu and spin  spaces and $\hat{G}^R $ is  the retarded Green function averaged over disorder. Using the standard diagram technique for averaging over the realizations of random potential~\cite{abrikosov_methods_1975} one gets
\begin{equation}\label{eq:Dyson}
\hat{G}^R_{0} (\bm{p},\epsilon)=\left(\epsilon_+  - H_0 (\bm{p}) -\hat{\Sigma}^R(\epsilon)  \right)^{-1},
\end{equation}
where $\epsilon_+ = \epsilon + i 0$,  $H_0 (\bm{p})$ is given by Eq.~\eqref{eq:HSC}, and the disorder-induced  self-energy is given by
\begin{align}
\label{eq:SelfEnergy1}
	\hat{\Sigma}^R(\epsilon) =\frac{-1}{2\pi\nu_{\mathrm n}\tau_{\mathrm{el}}}\int\frac{d^2p}{(2\pi)^2} \hat{G}^R(\epsilon, \bm{p}).
\end{align}
Integrating over momentum here, substituting the result into Eq.~\eqref{eq:Dyson}, and using Eq.~\eqref{eq:HSC} we write the retarded Green's function in the standard form
\begin{align}
	\hat G^R(\bm{p},\epsilon) = & \left(\tilde{\epsilon} - \left(\xi_{\bf p} - \bf b (\bf p) \cdot\bm\sigma\right) \tau_3 - \tilde\Delta \tau_1  \right. \nonumber \\
	 &+ \left.\bm h\cdot \bm \sigma + \bm v_\pm \cdot \bm p_{\mathrm s} \right)^{-1}.
\end{align}
where the disorder-renormalized energy $\tilde{\epsilon}$  and gap function $\tilde\Delta$ are given by the solutions of the  following equations:
\begin{subequations}
\label{eq:RenormEqs}
\begin{align}
	\tilde\epsilon &= \epsilon + \frac{i}{4\tau_{\mathrm{el}}}\sum_\pm \int\frac{d\phi}{2\pi} \frac{\tilde\epsilon - (v_F p_s \pm h)\cos\phi}{\sqrt{(\tilde\epsilon - (v_F p_s \pm h)\cos\phi)^2 -\tilde\Delta^2}} ,\\
	 \tilde\Delta &= \Delta+ \frac{i\tilde\Delta}{4\tau_{\mathrm{el}}}\sum_\pm \int\frac{d\phi}{2\pi} \frac{1}{\sqrt{(\tilde\epsilon - (v_F p_s \pm h)\cos\phi)^2 -\tilde\Delta^2}}.
\end{align}
\end{subequations}
Here, as in the ballistic case, we assumed longitudinal polarization, $\bm{p}_{\mathrm s} \parallel  {\bf p}^{\mathrm{(gs)}}_{\mathrm s}$.
Equations \ref{eq:RenormEqs} are similar to those arising in the theory of superconductors with magnetic impurities. It is easy to see
that in the absence of perturbations breaking time-reversal symmetry, ${\bf h}, {\bf p}_{\mathrm s}=0$, Eqs.~\eqref{eq:RenormEqs} yield $\frac{\tilde{\epsilon}}{\epsilon} = \frac{\tilde{\Delta}}{\Delta}$, which reproduces the standard BCS result for the density of states. For weak time-reversal breaking perturbations, $\left|(h \pm v_{\mathrm F} p_{\mathrm s}^{(\mathrm{gs})})/(\tilde\epsilon - \tilde\Delta)\right| \ll 1$ the density of states Eq.~\eqref{eq:DOS_GF} can be expressed in the form (see Appendix B for details)
\begin{subequations}
\label{eq:DiffusiveDOS}
\begin{align}
\label{eq:DiffusiveDOSa}
	\nu(\epsilon) &= \frac{\nu_{\mathrm n } A^{-1/3} \theta\left(\tilde w + \frac{3}{2^{5/3}}\right) }{\sqrt{3}}\left[\frac{\tilde\alpha(\tilde w)}{2^{4/3}}-\frac{2^{4/3}\tilde w^2}{\tilde\alpha(\tilde w)}\right],\\
	\tilde\alpha(\tilde w) &= \left(16 \tilde w^3 + 27 + 3\sqrt{3}\sqrt{32\tilde w^3+27}\right)^{1/3}.
\end{align}
\end{subequations}
 Here   $A =\left( h^2 + (v_{\mathrm F} p_{\mathrm s}^{(\mathrm{gs})})^2 \right) \tel \Delta$, and $\tilde w = w A^{-2/3}$. Using  Eq.~\eqref{eq:DiffusiveDOS} we evaluate the level sensitivity Eq.~\eqref{eq:LevelSensitivity2}, and substitute the result into~\eqref{eq:sigma_ratio} to obtain the Debye conductivity in the diffusive regime (see Appendix B)
\begin{eqnarray}
\label{eq:sigmadiffusivemu}
	\frac{\sigma_{\mathrm{DB}}}{\sigma_{\mathrm{n}}} & = &
	  \frac{ I_{\mathrm D} }{1+(\omega\tin)^2} \frac{\tau_{\mathrm{in}}}{\tau_{\mathrm{el}}}\frac{\Delta}{T} \left(\Delta \tau_{\mathrm{el}}^5 h^4\right)^{1/3} \nonumber \\
&&\times \frac{(\beta \cos\theta  -\alpha \sin \theta)^2 p_F^2}{4 E_F^2} \tilde g^2(T_c \tel)
\end{eqnarray}
Here  the numerical constant $I_{\mathrm D} \approx0.38727$ is given by a definite integral in Eq.~(B9) of the Appendix.

 In summary, we have identified a new mechanism of  magnetoconductivity of non-centrosymmetric superconductors, which arises from the quasiparticle spectral flow. It provides a contribution to the conductivity in the presence of an in-plane Zeeman field which is proportional to the inelastic quasiparticle relaxation time $\tau_{\mathrm{in}}$.  In the ballistic, $h \tau_{\mathrm{el}}^2\Delta \gg 1$, and diffusive, $h \tau_{\mathrm{el}}^2\Delta \ll 1$, regimes this contribution is described by Eqs.~\eqref{eq:sigma_ration_ballistic} and \eqref{eq:sigmadiffusivemu}, respectively.
Since under typical conditions $\tau_{\mathrm{in}}$ exceeds the elastic relaxation time $\tau_{\mathrm{el}}$ by several orders of magnitude this contribution may exceed  the conventional contribution proportional to $\tau_{\mathrm{el}}$ . Further, the Debye contribution to conductivity is strongly anisotropic; it exhibits a characteristic dependence on the angle between the direction of the external magnetic and the electric field of the microwave.  This dependence is different in the ballistic, \eqref{eq:sigma_ration_ballistic},  and diffusive, \eqref{eq:sigmadiffusivemu}, regimes. We also note that  in typical situations ,  the positive magneto-conductance turns out to be much larger than the $H_{\|}$-dependence of the conductivity in normal  metals, which may be estimated as
$
\frac{\sigma(H_{\|})-\sigma(0)}{\sigma(0)}\sim \left(\frac{\mu_{B} H_{\|}}{E_{F}}\right)^{2}
$.

Although we focused our consideration on the  interval of temperatures near $T_{c}$,  the mechanism of magnetoconductance discussed above is present
in a much broader temperature interval. In particular, even at small temperatures, $T\ll \Delta$, where  the quasiparticle concentration becomes exponentially small,  $\propto \exp(-\Delta/T)$ it may give a large contribution to the low frequency  magnetoconductivity.  In this regime the quasiparticle relaxation is characterized by two time scales:\emph{ i) } quasiparticle scattering  processes, which conserve the number of quasiparticles, occur on a time scale $\tau_{\mathrm{in,sc}}$, which is independent of the quasiparticle concentration, \emph{ii) }  quasiparticle recombination processes are characterized by a relaxation time, which is inversely proportional to the exponentially small  concentration of quasiparticles,  $\tau_{\mathrm{in,r}}\sim \tau_{\mathrm{in,sc}}\exp(\Delta/T)$. Since the Debye contribution to the conductivity is proportional to the longest relaxation time  in the system, at sufficiently low frequencies  the corresponding exponentially small factors cancel in the conductivity. Thus  $\sigma_{\mathrm{DB}}$ at low temperatures is, roughly speaking,  comparable to that near $T_c$. However, in  this case the frequency dispersion of $\sigma_{\mathrm{DB}}(\omega)$ takes place at very low frequencies, $\omega\sim 1/\tau_{\mathrm{in,r}}$.

\begin{acknowledgments}

The work of M.S. was supported by the National Science Foundation Grant MRSEC DMR-1719797. The work of A.A. was supported in part by the Russian Scientific Foundation under Grant No. 20-12-00361.
\end{acknowledgments}

\title{Giant magnetoconductivity in non-centrosymmetric superconductors appendix}

\author{M. Smith}

\affiliation{Department of Physics, University of Washington, Seattle, WA 98195 USA}

\author{A. V. Andreev} 

\affiliation{Skolkovo Institute of Science  and  Technology,  Moscow,  143026,  Russia}

\affiliation{Department of Physics, University of Washington, Seattle, WA 98195 USA}

\affiliation{L. D. Landau Institute for Theoretical Physics, Moscow, 119334 Russia}

\author{B. Z. Spivak}

\affiliation{Department of Physics, University of Washington, Seattle, WA 98195 USA}

\date{\today}

\maketitle

\addtolength{\abovedisplayskip}{-1mm}

\appendix

\section{Evaluation of the Ground State Supercurrent}
\label{sec:appendixA}
In this appendix we derive the expressions for the coefficients $a$ and $b$ in Eq.~(3) for the current in response to an in-plane magnetic field. In clean $T_c \tel \gg 1$ and dirty $T_c\tel \ll 1$ superconductors, they are given by Eqs.~(8) and~(9). In doing so we derive more general expressions for the in plane current due to an in-plane magnetic field, Eqs.~\eqref{eq:CurrentCleanSphericalFS} and~\eqref{eq:CurrentDirtySphericalFS}, which are valid for an arbitrary spectrum and weak spin-orbit coupling, $b \ll E_F$.
The clean case is treated in section~\ref{sec:Clean} and the dirty case is treated in section~\ref{sec:Dirty}.
The current in the ground state is written in terms of the single particle Green's function
\begin{align}
	\bm j = e T \sum_n \int \frac{d^2p}{(2\pi)^2} \Tr \left\{ \overline{\bm{\hat v}(\bm p) \hat G(i\epsilon_n, \bm p)} \right \}
\end{align}
where $e$ is the electric charge, $T$ is the temperature, $\epsilon_n = \pi T(2n+1)$ is a fermionic Matsubara frequency, $\bm{\hat v}(\bm p) = \frac{d}{d\bf p} \left(\xi_{\bf p} + \bm b(\bm p)\cdot\bm\sigma\right) $ is the velocity operator, $\Tr$ is a trace over Gorkov-Nambu space and spin space, and $\overline{...}$ denotes an average over random impurity positions. The Matsubara Green's function $\hat G = \left( i\epsilon_n - \hat H \right)^{-1}$ where $\hat H$ is
\begin{align}
	\hat H &= \hat H_0 + \hat H_1 \nonumber \\
	\hat H_0 &= (\xi_{\bf p} + \bm b(\bm p)\cdot \bm \sigma+ U(\bm r)) \tau_3  + \Delta\tau_1 \nonumber \\
	\hat H_1 &=  \bm{\hat v}(\bm p) \cdot \bps +\bm h \cdot \bm \sigma \nonumber.
\end{align}
 Below we treat $\hat H_1$ as a perturbation, and expand $\hat G$ in $\hat H_1$. As we are concerned with temperatures near $T_c$, we also expand in $\Delta$. The resulting expression for the current is 
\begin{align}
\label{eq:CurrentExpansion}
	\bm j &= e T\sum_n  \int \frac{d^2 p}{(2\pi)^2}\Tr \left \{ \overline{ \bm{\hat{v}} \hat G_n \Delta\tau_1 \hat G_n \hat{H}_1 \hat G_n \Delta\tau_1 \hat G_n}  \right.  \nonumber\\
	&\left. + \overline{ \bm{\hat{v}}\left(\hat G_n \hat{H}_1 \hat G_n \Delta \tau_1 \hat G_n \Delta\tau_1 \hat G_n + \hat G_n \Delta\tau_1 \hat G_n \Delta\tau_1 \hat G_n \hat H_1 \hat G_n \right)}\right\}
\end{align}

\subsection{Clean Regime $T_c \tel \gg 1$}
\label{sec:Clean}
\begin{figure}[h]
	\begin{subfigure}{.225\textwidth}
		\centering
		\includegraphics[width=4cm]{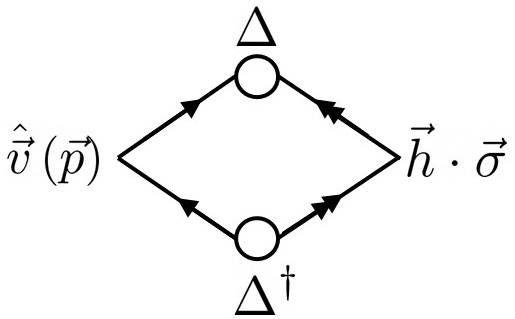}
		\caption{Figure 1a}
	\end{subfigure}
	\begin{subfigure}{.225\textwidth}
		\centering
		\includegraphics[width=4cm]{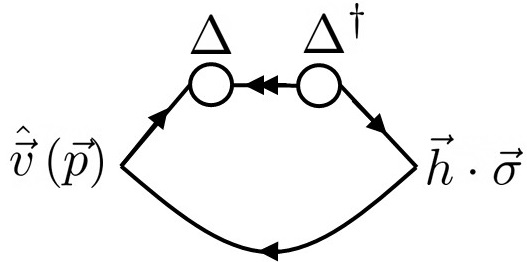}
		\caption{Figure 1b}
	\end{subfigure}
	\begin{subfigure}{.5\textwidth}
		\centering
		\includegraphics[width=4cm]{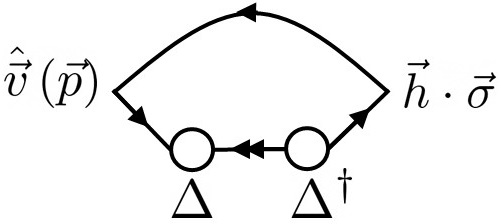}
		\caption{Figure 1c}
	\end{subfigure}
	\caption{The diagrams contributing to the current due to an in-plane Zeeman field in the clean limit $\tel T_c \gg 1$. Lines with single arrows denote $G_n(i\epsilon_n, \bm p)$ and lines with double arrows denote $-G_n^\dagger(i\epsilon_n,\bm p)$.}
	\label{fig:Fig1}
\end{figure}
In the clean limit, $T_c\tel \gg 1$, impurity scattering can be neglected and the Matsubara Green's functions in Eq.~\eqref{eq:CurrentExpansion} are their given by their values in the absence of an impurity potential,
\begin{subequations}
\begin{align}
	\hat G_n &= \left(\begin{array}{cc}
	\widecheck G(i\epsilon_n, \bm p) & 0\\
	0 & -\widecheck G^\dagger (i\epsilon_n, \bm p)
	\end{array}\right) \\
\label{eq:GF}
	\widecheck G &= \left( i\epsilon_n - \xi - \bm b\cdot \bm \sigma\right)^{-1} 
\end{align}
\end{subequations}

The diagrams in Fig.~\ref{fig:Fig1} correspond to the three terms in Eq.~\eqref{eq:CurrentExpansion} at zero superfluid momentum. Evaluation of these diagrams gives the following expression for the current due to an in-plane Zeeman field
\begin{align}
\label{eq:CurrentCleanSphericalFS}
	\bm j_\mu^c &= 2e \Delta^2 T\pi \sum_n \left[\frac{\nu_0 \left\langle \frac{d b_0^\alpha}{d \bf p_{\mathrm F}} h^\alpha \right\rangle}{|\epsilon_n|(\epsilon_n^2 + b_0^2)} + \frac{\nu_0 \left\langle \frac{d b_0^2}{d\bf p_{\mathrm F}}  \bm h \cdot \bm b_0 \right \rangle }{2|\epsilon_n|^3(\epsilon_n^2 + b_0^2)}\right. \nonumber\\ 
	 &- \left.\frac{\left\langle \nu_0\left( \bm v_0 \bm b_1\cdot \bm h + \bm v_1 \bm b_0\cdot\bm h\right) + \nu_1 \bm v_0 \bm b_0 \cdot \bm h\ \right \rangle}{|\epsilon_n|^3}\right].
\end{align}
Here $\bm v_0 = \bm v |_{\xi = 0}$ is the Fermi velocity, $\bm v_1 = d \bm v/d\xi |_{\xi = 0}$, $\bm b_0 = \bm b|_{\xi = 0}$ is the spin-orbit coupling on the Fermi surface, $\bm b_1 = d\bm b/d\xi |_{\xi = 0}$, $\nu_0 = m/2\pi$ is  the density of states for a single spin polarization to lowest order in $b_0$, and $\nu_1 = d\nu(\xi)/d\xi|_{\xi=0}$. Further $\langle ... \rangle = \int_0^{2\pi} \frac{d\phi}{2\pi} ... $ denotes an average over the Fermi circle. 

In obtaining Eq.~\eqref{eq:CurrentCleanSphericalFS} it is convenient to express the Green's function Eq.~\eqref{eq:GF} in terms of projectors $\Pi^\rho$ onto the $\rho$ Fermi surface,
\begin{subequations}
\begin{align}
	G (i\epsilon_n, \bm p ) = \sum_{\rho = \pm 1} \widecheck \Pi^\rho\ G_\rho(i\epsilon_n,\bm p)\\
	\widecheck \Pi^\rho = \frac{1}{2}(1 + \rho \hat b(\bm p) \cdot \bm \sigma)\\
	G_\rho(i\epsilon_n, \bm p) = \left( i \epsilon_n - \xi_{\bf p} - \rho b(\bm p) \right)^{-1}.
\end{align}
\end{subequations}
One can reduce the problem of considering two Fermi surfaces to the consideration of a single Fermi surface as in Eq.~\eqref{eq:CurrentCleanSphericalFS} by expanding the spin orbit-coupling about $\xi_{\bf p} = 0$. For linear spin-orbit coupling, such as considered in Refs.~\cite{edelstein_magnetoelectric_1995,edelstein_magnetoelectric_2005} this approximation is exact. 

So far we have made no assumptions about the spectrum, only assuming that spin-orbit coupling is weak in comparison to the Fermi energy $b_0 \ll E_F$. For the spectrum assumed in the main text, with spin-orbit coupling given by Eq.~(5), $\nu_1 = 0$. Equation~\eqref{eq:CurrentCleanSphericalFS} then reduces to

\begin{align}
\label{eq:CleanCurrentSimple}
	\bm j_\mu^{c} = - \frac{ e p_{\mathrm F}^2}{2 E_{\mathrm F}} \Delta^2 \sum_n \frac{b_0^2  \left[ \alpha \bm n \times \bm h + \beta \bm h\right]}{2|\epsilon_n|^3 \left(\epsilon_n^2 + b_0^2\right)} p_{\mathrm F}
\end{align}
Eq.~\eqref{eq:CleanCurrentSimple} reduces to Eq.~(8) in the main text for the $T_c\tel \gg 1$ case by taking $b_0 \gg T_c$. Eq.~\eqref{eq:CleanCurrentSimple} reproduces the result of Edelstein for Rashba-type spin-orbit coupling ($\beta = 0$)~\cite{edelstein_magnetoelectric_1995}.

\subsection{Dirty Regime $T_c \tel \ll 1$}
\label{sec:Dirty}
We now account for impurity scattering in the regime $ T_{\mathrm c} \tel \ll 1$. Because of the presence of Green's functions with opposite frequencies impurity averaging does not reduce to simply replacing the Green's functions by their disorder-averaged values. The averaging of products of Green's functions leads to the appearance impurity ladders, as shown in Figs~\ref{fig:Fig4a} - \ref{fig:Fig4d}. The diagrams~\ref{fig:Fig4a} - \ref{fig:Fig4d} give the dominant contribution to the ground state current in the small parameter $T_c\tel$. 

In the presence of impurities the quasiparticle energy $\epsilon_n \rightarrow \tilde\epsilon_n = \epsilon_n + \frac{1}{2\tel} \sgn \epsilon_n$, where $\tel^{-1} = m n_{\mathrm i} u^2$ is the elastic scattering rate, $n_{\mathrm i}$ is the impurity concentration, $m$ is the electron mass, and $u$ is the amplitude of the impurity potential in momentum space. In particular, the order parameter vertex is renormalized by the impurity ladder.

\subsubsection{Renormalization of Order Parameter}
\begin{figure}[H]
	\centering
	\includegraphics[width=6cm]{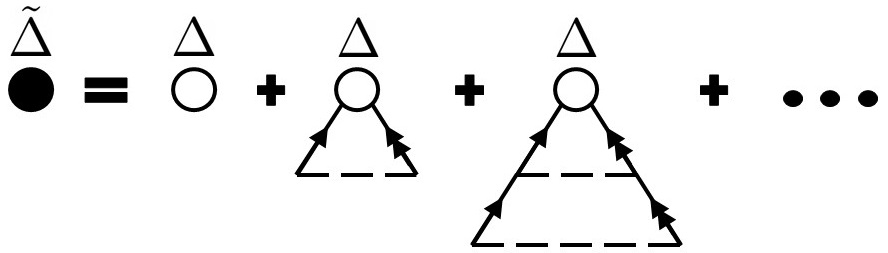}
	\caption{The renormalization of the order parameter due to impurity scattering.  Lines with single arrows denote $G_n(i\tilde\epsilon_n, \bm p)$ and lines with double arrows denote $-G_n^\dagger(i\tilde\epsilon_n,\bm p)$. Dashed impurity lines correspond to a factor $n_{\mathrm i} u^2$.}
	\label{fig:Fig2}
\end{figure}
The renormalization of the order parameter is shown diagramatically in Fig.~\ref{fig:Fig2}. It is given by
\begin{align}
	\tilde\Delta &= \Delta + n_i u^2 \tilde\Delta \int \frac{d^2 p}{(2\pi)^2} \hatGd \hatGdagd \nonumber\\
	\label{eq:OPRenorm2}
	 &= \Delta\left(1 + \frac{1}{2\tel |\epsilon_n|}\right)
\end{align}


\subsubsection{Diffuson Ladder}
\begin{figure}[H]
	\centering
	\includegraphics[width=8cm]{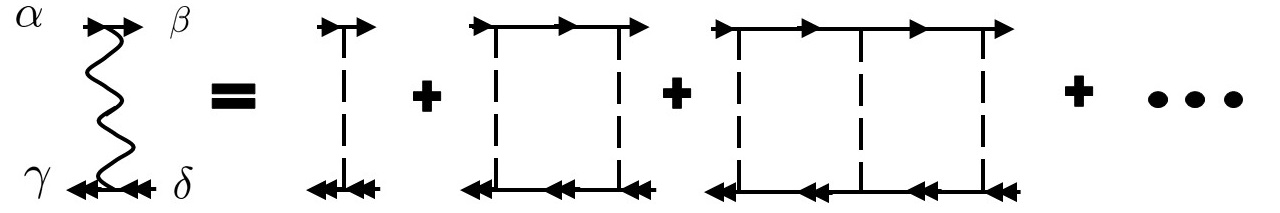}
	\caption{The diffuson ladder that appears in diagrams~\ref{fig:Fig4a}~-~\ref{fig:Fig4d}. Lines with single arrows denote $G_n(i\tilde\epsilon_n, \bm p)$ and lines with double arrows denote $-G_n^\dagger(i\tilde\epsilon_n,\bm p)$. Dashed impurity lines correspond to a factor $n_{\mathrm i} u^2$.}
	\label{fig:Fig3}
\end{figure}
In the presence of spin-orbit coupling the impurity ladder acquires a non-trivial dependence on the spin indices. Evaluation of the diagram Fig~\ref{fig:Fig3} yields
\begin{align}
\label{eq:T}
	 &T^{\alpha \beta}_{\gamma \delta} = - n_{\mathrm i} u^2 \delta^{\alpha \beta} \delta^{\gamma\delta} \nonumber \\
&+ n_{\mathrm i} u^2 \int \frac{d^2 p}{(2\pi)^2} \underline T^{\alpha \alpha_1}_{\gamma\gamma_1} \hat G_{\alpha_1 \beta}(i\tilde\epsilon_n,\bm p) \hat G^\dagger_{\delta \gamma_1} (i\tilde\epsilon_n,\bm p)
\end{align}

We assume $\left\langle \hat b^a \hat b^b \right\rangle = \delta_2^{ab}/2$, where $\delta_2^{ab}$ is the two-dimensional Kronecker delta. Evaluation of Eq.~\eqref{eq:T} gives

\begin{align}
\label{eq:TFinal}
	T^{\alpha\beta}_{\gamma\delta} &= \sum_{i=0}^3 A_i \sigma_i^{\alpha\gamma} \sigma_i^{\delta\beta}  =-\frac{1}{4\pi\tel \nu_0} \left[ \frac{\delta^{\alpha\gamma}\delta^{\delta\beta}}{\tilde Z_1(\tilde\epsilon_n) + \tilde Z_2(\tilde\epsilon_n)} \right. \nonumber\\
	&+\left. \frac{\sigma_3^{\alpha\gamma}\sigma_3^{\delta\beta}}{\tilde Z_1(\tilde\epsilon_n) - \tilde Z_2(\tilde\epsilon_n)} + \frac{\sigma_1^{\alpha\gamma}\sigma_1^{\delta\beta}}{\tilde Z_1(\tilde\epsilon_n) } + \frac{\sigma_2^{\alpha\gamma}\sigma_2^{\delta\beta}}{\tilde Z_1(\tilde\epsilon_n) }\right]
\end{align}

where
\begin{subequations}
\begin{align}
	\tilde Z_1(\tilde\epsilon_n) &= 1- \frac{1}{4\tel \tilde\epsilon_n|} \left( 1+ \frac{\tilde\epsilon_n^2}{\tilde\epsilon_n^2 + b^2}\right)\\
	\tilde Z_2(\tilde\epsilon_n) &=-\frac{1}{4\tel |\tilde\epsilon_n|}\frac{b^2}{\tilde\epsilon_n^2 + b^2}
\end{align}
\end{subequations}

\subsubsection{Diagrams Contributing to the Ground State Current when $T_c \tel \ll 1$}
\begin{figure}[h!]
	\begin{subfigure}{.225\textwidth}
		\centering
		\includegraphics[width=4cm]{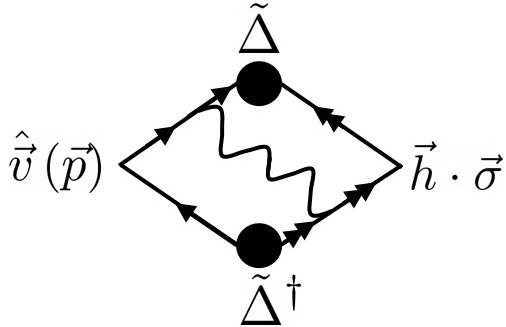}
		\caption{Figure 4a}
		\label{fig:Fig4a}
	\end{subfigure}
	\begin{subfigure}{.225\textwidth}
		\centering
		\includegraphics[width=4cm]{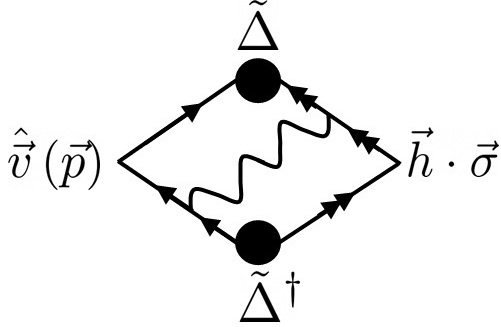}
		\caption{Figure 4b}
		\label{fig:Fig4b}
	\end{subfigure}
	\begin{subfigure}{.25\textwidth}
		\centering
		\includegraphics[width=4cm]{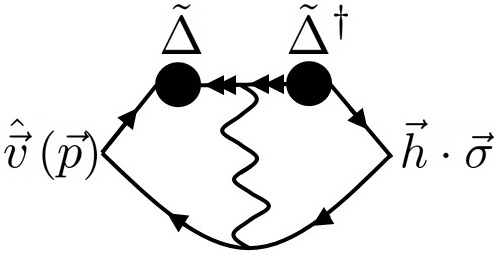}
		\caption{Figure 4c}
		\label{fig:Fig4c}
	\end{subfigure}
	\begin{subfigure}{.225\textwidth}
		\centering
		\includegraphics[width=4cm]{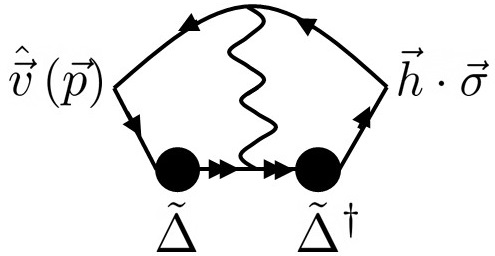}
		\caption{Figure 4d}
		\label{fig:Fig4d}
	\end{subfigure}
		\caption{The diagrams giving the dominant contribution to the current due to an in-plane magnetic field in the dirty regime $T_c\tel \ll 1$. Lines with single arrows denote $G_n(i\tilde\epsilon_n, \bm p)$ and lines with double arrows denote $-G_n^\dagger(i\tilde\epsilon_n,\bm p)$.}
		\label{fig:Fig4}
\end{figure}

\begin{figure}[h!]
	\begin{subfigure}{.225\textwidth}
		\centering
		\includegraphics[width=4cm]{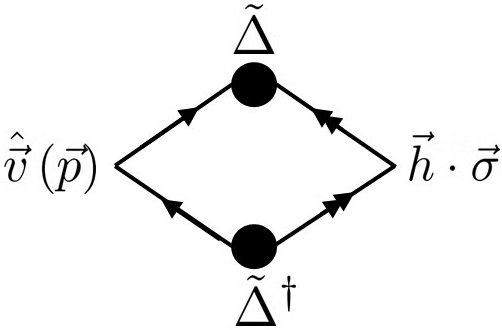}
		\caption{Figure 5a}
		\label{fig:Fig5a}
	\end{subfigure}
	\begin{subfigure}{.2\textwidth}
		\centering
		\includegraphics[width=4cm]{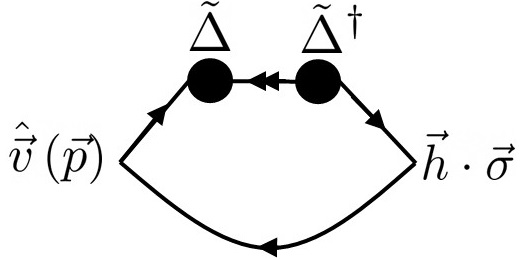}
		\caption{Figure 5b}
		\label{fig:Fig5b}
	\end{subfigure}
	\begin{subfigure}{.5\textwidth}
		\centering
		\includegraphics[width=4cm]{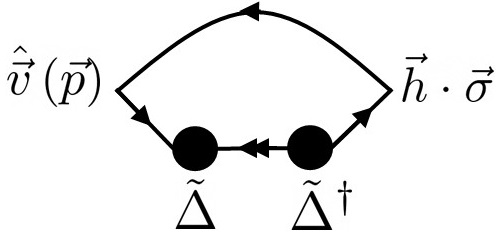}
		\caption{Figure 5c}
		\label{fig:Fig5c}
	\end{subfigure}
	\caption{The diagrams that give sub-leading contributions to the current in the dirty regime $T_c\tel \ll 1$. Lines with single arrows denote $G_n(i\tilde\epsilon_n, \bm p)$ and lines with double arrows denote $-G_n^\dagger(i\tilde\epsilon_n,\bm p)$.}
	\label{fig:Fig5}
\end{figure}

The contribution to the current from diagrams~\ref{fig:Fig5a} - \ref{fig:Fig5c}
are subleading to the contribution to the current arising from diagrams~\ref{fig:Fig4a} - \ref{fig:Fig4d} in the small parameter $T_c \tel$.  The contribution from the latter diagrams, while algebraicly laborious, can written and simplified in a similar fashion to section~\ref{sec:Clean}. Assuming $T_c \tel \ll 1$ and that the Fermi circle and spectrum are isotropic, we obtain for the ground state current induced by an in-plane Zeeman field
\begin{widetext}
\begin{align}
\label{eq:CurrentDirtySphericalFS}
	\bm j^d_\mu &=  -4eT \sum_n \frac{\tilde\Delta^2 A_1 \nu_0\pi^2}{\tilde\epsilon_n^2 + b_0^2}\left[-\frac{ \left\langle \nu_0\left( \bm v_0(p_{\mathrm F}) \bm b_1(p_{\mathrm F}) \cdot \bm h + \bm v_1(p_{\mathrm F}) \bm b_0(p_{\mathrm F}) \cdot \bm h\right) + \nu_1 \bm v_0(p_{\mathrm F}) \bm b_0(p_{\mathrm F}) \cdot \bm h\right\rangle}{2\tilde\epsilon_n^2}  + \frac{\nu_0 \left\langle  h^\alpha \frac{d b_0^\alpha(p_{\mathrm F})}{d \bf p_{\mathrm F} }\right\rangle}{2\left(\tilde\epsilon_n^2+b_0^2\right)}\right. \nonumber\\
    &+\left. \frac{\nu_0  \left \langle  \frac{d b_0^\alpha(p_{\mathrm F})}{d \bf p_{\mathrm F}} b_0^\alpha(p_{\mathrm F}) \bm h\cdot \bm b_0(p_{\mathrm F})\right\rangle}{2\tilde\epsilon_n^4(\tilde\epsilon_n^2+b_0^2)}   +\frac{\nu_0  \left\langle  \frac{d b_0^\alpha(p_{\mathrm F})}{d \bf p_{\mathrm F} } b_0^\alpha(p_{\mathrm F}) \bm b_0(p'_{\mathrm F}) \cdot \bm b(p_{\mathrm F}) \bm h \cdot \bm b(p'_{\mathrm F})\right\rangle}{\tilde\epsilon_n^4(\tilde\epsilon_n^2 + b_0^2)} + \frac{\nu_0 \left \langle \bm h\cdot \bm b_0(p'_{\mathrm F}) b^\alpha_0(p'_{\mathrm F})  \frac{d b_0^\alpha(p_{\mathrm F})}{d \bf p_{\mathrm F} }\right\rangle}{\tilde\epsilon_n^2 (\tilde\epsilon_n^2+b_0^2)} \right. \nonumber\\
    &- \left.    \frac{\left\langle \bm h\cdot \bm b_0(p'_{\mathrm F}) \left[ \nu_0 \left( \bm v_0(p_{\mathrm F}) \bm b_1(p_{\mathrm F}) \cdot \bm b_0(p'_{\mathrm F}) + \bm v_1(p_{\mathrm F}) \bm b_0(p_{\mathrm F})\cdot \bm b_0(p'_{\mathrm F})\right) + \nu_1 \bm v_0 \bm b_0(p_{\mathrm F}) \cdot \bm b_0(p'_{\mathrm F})\right]\right\rangle}{\tilde\epsilon_n^4}   \right] 
\end{align}
\end{widetext}
where $\langle ... \rangle = \int \frac{d\phi_{\bf p_{\mathrm F}} d\phi_{\bf p'_{\mathrm F}}}{(2\pi)^2} ...\ $. For the spectrum assumed in the main text, with the spin-orbit coupling given by Eq.~(5),
$\nu_1 = 0$ as in Sec.~\ref{sec:Clean}. Equation~\eqref{eq:CurrentDirtySphericalFS} reduces to
\begin{align}
\label{eq:CurrentDirty}
	\bm j^d_\mu &= - \frac{e p_{\mathrm F}^2}{2 E_{\mathrm F}} \left(\Delta\tel\right)^2 T\sum_n \frac{b_0^2\left[ \alpha \bm n \times \bm h + \beta \bm h \right]}{|\epsilon_n| + b_0^2 \tel} 
\end{align}
Eq.~\eqref{eq:CurrentDirty} reduces to Eq.~(8) for the case $T_c\tel \ll 1$ if we assume $b_0 \gg \sqrt{T_c \tel^{-1}}$.
\section{Evaluation of the Density of States and Debye Conductivity in the Diffusive Regime $h\tel^2\Delta \ll 1$}
\label{sec:appendixB}

To derive Eqs. (27) and (28) in the main text we begin by writing the equations (26) for the renormalized energy $\tilde\epsilon$ and renormalized order parameter $\tilde\Delta$ in terms of the variables $z = (\tilde\epsilon - \tilde\Delta)/\tilde\Delta$ and $d = \Delta/\tilde\Delta$. Assuming the magnetic field to be weak we focus on the energy interval near the gap and introduce small parameters parameters $\gamma_\pm = \left(v_F p_s^{(gs)} \pm h\right)/\Delta$, $w = (\epsilon-\Delta)/\Delta$, and $\beta = (\tel\Delta)^{-1}$, and rewrite Eq. (26) in the form,
\begin{subequations}
\label{eq:RenormEqs1}
\begin{align}
	\label{eq:RenormEqs1a}
	 \frac{1+z}{d} &= 1+w + \frac{i\beta}{4}\sum_\pm \int\frac{d\phi}{2\pi} \frac{1+z -d\gamma_\pm\cos\phi}{\sqrt{(1+z-d\gamma_\pm\cos\phi)^2-1}},\\
	 \label{eq:RenormEqs1b}
	 \frac{1}{d} &= 1+ \frac{i\beta}{4}\sum_\pm \int\frac{d\phi}{2\pi} \frac{1}{\sqrt{(1+z-d\gamma_\pm\cos\phi)^2-1}}.
\end{align}
\end{subequations}

Performing the angular integration in Eq.~\eqref{eq:RenormEqs1} gives lengthy expressions involving elliptic integrals of the first and second kind. In the absence of time-reveral breaking terms, $\gamma_\pm = 0$, we recover the BCS result  $\tilde \epsilon/\epsilon = \tilde\Delta/\Delta$, or in our variables $z = w$. The corrections to $z$ due to $\gamma_\pm$ being finite are small. Further, the energy interval that contributes to the Debye conductivity is when $|w| \ll 1$. Thus, we expand Eqs.~\eqref{eq:RenormEqs1} in the small parameter $\left|d \gamma_\pm/z\right| = |(h \pm v_{\mathrm F} p_{\mathrm s}^{(\mathrm{gs})})/(\tilde\epsilon - \tilde\Delta)|\ll 1$, and use Eq.~\eqref{eq:RenormEqs1b} to simplify Eq.~\eqref{eq:RenormEqs1a} to obtain
\begin{subequations}
\label{eq:RenormEqs2}
\begin{align}
	\label{eq:RenormEqs2a}
	\frac{z}{d} &= w + \frac{i\beta}{8\sqrt{2}} \left[ 4 \sqrt{z} - \frac{1}{4}\frac{d^2 \gamma^2}{z^{3/2}}\right],\\
	\label{eq:RenormEqs2b}
	\frac{1}{d} &= 1 + \frac{i\beta}{8 \sqrt{2}} \left[ \frac{4}{\sqrt{z}} - \frac{3}{4} \frac{d^2 \gamma^2}{z^{5/2}} \right],
\end{align}
\end{subequations}
where $\gamma^2 = \left(h^2 + \left(v_{\mathrm F} p_{\mathrm s}^{(\mathrm{gs})}\right)^2 \right)/\Delta^2$. As we are interested in only the leading order correction in  $\gamma$ to these equations, we substitute Eq.~\eqref{eq:RenormEqs2b} into Eq.~\eqref{eq:RenormEqs2a} and keep only the lowest order terms in $\gamma$.  We also make the substitution $y = z^{-1/2}$, and obtain a cubic equation for $y$

\begin{align}
\label{eq:Cubic}
	\frac{i\gamma^2}{2\sqrt{2}\beta}y^3 - y^2 w +1 = 0.
\end{align}
From Eq.~\eqref{eq:Cubic} at $w = 0$ we see $z \sim \gamma^{4/3} \beta^{-2/3}$. Using this and Eq.~\eqref{eq:RenormEqs2b} $d \sim \gamma^{2/3}\beta^{-4/3}$. Thus Eq.~\eqref{eq:Cubic} is only valid when $h\tel^2\Delta \ll 1$.

We do not write the second independent equation for $d$ as we can express the density of states in terms of only $y$. The  density of states is written in terms of the single particle Green's function, Eq.~(22). It is written in terms of variables $z$ and $d$ as
\begin{equation}
	 \nu(\epsilon) = \frac{\nu_n}{2}\sum_{\pm}\Re \int_0^{2
\pi}\frac{d\phi}{2\pi} \frac{1+z-d\gamma_\pm\cos\phi}{\sqrt{(1+z-d\gamma_\pm\cos\phi)^2 - 1}}.
\end{equation}
Making the same approximations as we did for Eqs.~\eqref{eq:RenormEqs1}, and using the variable $y$, we find 
\begin{align}
	\nu(\epsilon) = \frac{\nu_n}{\sqrt{2}}\Re y.
\end{align}
Thus, to obtain the density of states we take the correct root of Eq.~\eqref{eq:Cubic} such that $\Re y \geq 0$. The density of states is then given by Eqs.~(27) in the main text. Substituting Eq.~(27a) into  Eq.~(14) for the level sensitivity, we obtain

\begin{equation}
\label{eq:DirtyLevelSensitivity}
	V(\epsilon) = -v_{\mathrm F} \frac{v_{\mathrm F} p_{\mathrm s}^{(\mathrm{gs})}}{\Delta}\left(\frac{\gamma}{\beta}\right)^{2/3}\tilde V(\tilde w)
\end{equation}
where $\tilde V$ is given by
\begin{widetext}
\begin{equation}
	\tilde V(\tilde w) = \frac{2}{\tilde\nu(\tilde w)}\int_{-3/2^{5/3}}^{\tilde w}d\tilde w' \frac{1}{\sqrt{3}}\left[\frac{2^{4/3}\tilde w^2}{\tilde{\mathrm\alpha}(\tilde w)}-\frac{\tilde{\mathrm\alpha}(\tilde w)}{2^{4/3}}+\left(\frac{1}{2^{4/3}}+\frac{2^{4/3}\tilde w^2}{\tilde{\mathrm\alpha}^2(\tilde w)}\right)\frac{2}{3\tilde{\mathrm\alpha}^2(\tilde w)}\left[27+\sqrt{27}\frac{16\tilde w^3 + 27}{\sqrt{32\tilde w^3+27}} \right] \right]
\end{equation}
\end{widetext}
and  $\tilde\nu$ is
\begin{subequations}
\begin{eqnarray}
	\tilde\nu(\epsilon) &=&  \theta\left(\tilde w + \frac{3}{2^{5/3}}\right) \frac{1}{\sqrt{3}}\left[\frac{\tilde\alpha(\tilde w)}{2^{4/3}}-\frac{2^{4/3}\tilde w^2}{\tilde\alpha(\tilde w)}\right],\\
	\tilde\alpha(\tilde w) &=& \left(16 \tilde w^3 + 27 + 3\sqrt{3}\sqrt{32\tilde w^3+27}\right)^{1/3}.
\end{eqnarray}
\end{subequations}
Then, substituting Eq.~(27a) and~\eqref{eq:DirtyLevelSensitivity} into Eq.~(13) we obtain Eq.~(28) for the Debye conductivity in the diffusive regime, with $I_{\mathrm D}$ given by
\begin{equation}
\label{eq:Id}
	I_{\mathrm D} = \frac{3}{4}\int_{-3/2^{5/3}}^\infty d\tilde w\ \tilde\nu(\tilde w)\ \tilde V^2(\tilde w) \approx 0.38727
\end{equation}

\end{document}